\input ./style/arxiv-general.cfg
\documentclass[aoas,MSNbibl,nameyear,dvips]{arximspdf}
\makeatletter
   \@ifpackageloaded{graphicx}{}{\usepackage{graphicx}}
\makeatother
\usepackage{multirow}

\doi{10.1214/15-AOAS828}% Updated by VTEXPTS2LaTeX.exe, 14.05.2015
\volume{9}
\issue{2}
\pubyear{2015}
\firstpage{754}
\lastpage{776}
\docsubty{FLA}

\makeatletter

\def\sklfrac#1#2{(#1/#2)}

\newcommand{\rrVert}{\Vert}
\newcommand{\llVert}{\Vert}
\renewcommand{\mid}{|}
\def\real{\mathbb R} % real numbers
\newcommand{\fbold}{\mathbf{f}}
\newcommand{\gbold}{\mathbf{g}}
\newcommand{\hbold}{\mathbf{h}}
\newcommand{\xbold}{\mathbf{x}}
\newcommand{\ybold}{\mathbf{y}}
\newcommand{\Wbold}{\mathbf{W}}
\newcommand{\etabold}{\bolds{\eta}}
\newcommand{\epsilonbold}{\bolds{\varepsilon}}
\newcommand{\thetabold}{\bolds{\theta}}
\makeatother

\begin{document}
\begin{frontmatter}

\title{Goodness of fit in nonlinear dynamics: Misspecified rates or misspecified states?}
\runtitle{Goodness of fit in nonlinear dynamics}

\begin{aug}
% Corresponding author: Giles Hooker - gjh27@cornell.edu% Updated by
%VTEXPTS2LaTeX.exe, 14.05.2015 10:05
\author[A]{\fnms{Giles}~\snm{Hooker}\corref{}\ead[label=e1]{gjh27@cornell.edu}\thanksref{T1}}
\and
\author[B]{\fnms{Stephen P.}~\snm{Ellner}\thanksref{T2}\ead[label=e2]{spe2@cornell.edu}}
\runauthor{G. Hooker and S.~P. Ellner}
\affiliation{Cornell University}
%\dedicated{}
\address[A]{Department of Biological Statistics\\
\quad and Computational Biology\\
Cornell University\\
Ithaca, New York 14853-4201\\
USA\\
\printead{e1}}
\address[B]{Department of Ecology\\
\quad and Evolutionary Biology\\
Cornell University\\
Ithaca, New York 14853-4201\\
USA\\
\printead{e2}}
\end{aug}
\thankstext{T1}{Supported in part by NSF Grants DEB-1353039 and DMS-10-53252.
This work was partly carried out while visiting the Department of
Mathematics and Statistics at the University of Melbourne.}
\thankstext{T2}{Supported in part by NSF Grants DEB-0813743 and DEB-125619.}

% HISTORY:
%
\received{\smonth{12} \syear{2013}}% Updated by VTEXPTS2LaTeX.exe,
%14.05.2015 10:05
%
\revised{\smonth{12} \syear{2014}}% Updated by VTEXPTS2LaTeX.exe,
%14.05.2015 10:05

% ABSTRACT
%
\begin{abstract}
This paper introduces diagnostic tests for the nature of lack of fit in
ordinary differential
equation models (ODEs) proposed for data. We present a hierarchy of
three possible sources
of lack of fit: unaccounted-for stochastic variation, misspecification
of functional forms
in rate equations, and omission of dynamic variables in the description
of the system.
We represent lack of fit by allowing a parameter vector to vary over
time, and propose generic
testing procedures that do not rely on specific alternative models.
Instead, different sources
for lack of fit are characterized in terms of nonparametric
relationships among latent variables.
The tests are carried out through a combination of residual bootstrap
and permutation methods.
We demonstrate the effectiveness of these tests on simulated data and
on real data
from laboratory ecological experiments and electro-cardiogram data.
\end{abstract}

% KEYWORDS
\begin{keyword}
\kwd{Differential equation}
\kwd{diagnostics}
\kwd{goodness of fit}
\kwd{attractor reconstruction}
\kwd{bootstrap}
\end{keyword}
\end{frontmatter}

%s1 #&#
\section{Introduction}\label{sec1}

Recent statistical literature has seen substantial interest in the
problem of fitting nonlinear continuous-time dynamical system models to
data. Statistical problems include estimating parameters, determining
parameter identifiability, experimental design, and testing goodness of
fit. These topics have been approached from numerous perspectives and
using various models, from deterministic models in the form of ordinary
differential equations (ODEs) through stochastic models based on Wiener
processes or finite population models such as branching processes.
Techniques for fitting models include nonlinear least squares
[\citet
{bock83,BatesWatts88,arorabiegler,GCC10}], maximizing likelihoods for
stochastic systems through particle filters [\citet{ionides06}] or via
equivalent Bayesian methods [e.g., \citet{golightly2011}], methods based
on pre-smoothing [\citet{BellmanRoth71,varah82,Ellner2002,Wu2012a}],
mimicking forecast models [\citet{PascualEllner00}] or indirect
inference [\citet{GourierouxMonfort96}], and fitting summary statistics
[\citet{TienGuckenheimer08,Ratmann2009,Reuman2006a,Wood2010a}].
\citet{RamsayDE} combine the criteria from least squares and from
pre-smoothing methods to achieve the advantages of each.

This paper presents an approach to model diagnostics for improving the
fit of a dynamical systems model. \citet{Hooker08} proposed a
goodness-of-fit test for ODE models using a likelihood ratio test. Here
we assume that a proposed ODE model has been found to fit poorly, so
the next goal is to distinguish among different potential sources of
model misspecification. In particular, we suppose that the proposed
model is an ODE
%e2 #&#
%
\begin{equation}
\label{ode} \frac{d}{dt} \xbold= \fbold(\xbold;t,\thetabold)
\end{equation}
in which $\xbold\in\real^d$ describes the state of the system and
$\fbold(\xbold;t,\thetabold)$ describes how quickly the system
changes at location $\xbold$ in the state-space,
depending on a vector of model parameters $\thetabold$ to be
estimated. We assume that we have vector-valued data $\ybold_1,\ldots
,\ybold_n$ from this system observed at times $t_1,\ldots,t_n$, where
$\ybold_i$ is related to $\xbold(t_i)$ by a known, possibly indirect,
measurement process. If we find that the model cannot fit the data
well, we then wish to improve the fit by changing the model in some
way. Here, we develop testing methods to distinguish between three
likely reasons for lack of fit, which would imply three different
directions for improving the model:
\begin{longlist}[1.]
\item[1.] Unmodeled disturbances unrelated to system dynamics, which if
modeled as random suggests a probabilistic description of system dynamics.

\item[2.] Misspecification of the parametric form of $\fbold$.

\item[3.] Misspecification of the state vector $\xbold$, in particular,
that the state vector $\xbold$ omits some variables that are needed to
provide a full description of the system state.
\end{longlist}
The methods we propose can be used in combination with a variety of
methods for parameter estimation in ordinary differential equations, as
discussed below. The same ideas can be employed for model improvement
in stochastic systems which propose a probabilistic model for the
evolution of $\xbold$. However, applications to stochastic systems
will require modifications to some of the details below and we will not
examine these further.

\citet{Hooker08} notes that
residuals from solutions to differential equation models give poor graphical
indications of how lack of fit should be addressed. This is because the
models describe the derivatives $d\xbold/dt$ rather than the
(observed) state
variables themselves. Instead, \citet{Hooker08} proposed
estimating lack of fit in terms of
\textit{empirical forcing functions}. These are nonparametric functions
$\gbold(t)$ which modify (\ref{ode}) to
%e3 #&#
%
\begin{equation}
\label{forcing} \frac{d}{dt} \xbold(t) = \fbold\bigl(\xbold
(t);t,\thetabold
\bigr) + \gbold(t)
\end{equation}
in such a way that a good fit to the data is achieved.
$\gbold(t)$
will thus represent both
random disturbances to the system and deterministic lack of fit in
$\fbold$.

The estimated $\gbold(t)$ can now be examined graphically
by plotting its relationship to $\xbold(t)$, along with lagged values
of both $\xbold$ and $\gbold$, although this can only be done
comprehensively when $\xbold$ is relatively low dimensional. In ODE
models, local (in time or state-space)
disturbances to the system are usually modeled as affecting $d\xbold
/dt$. These modify future values of $\xbold$,
so the effects of the disturbances will persist over time in the
observations. However, they can be accounted for locally
in $\gbold$. \citet{Hooker08} provides approximate
goodness-of-fit tests for the null hypothesis $\gbold\equiv0$
based on a basis expansion, $\gbold= \Psi(t) D$ for a vector of basis
functions $\Psi(t)$, and a
coefficient matrix $D$.

In this paper, we take the same approach, but we model lack of fit in a
more general way that includes
the possibility of parameter values changing over time, producing the system
%e4 #&#
%
\begin{equation}
\label{forcing2} \frac{d}{dt} \xbold(t) = \fbold\bigl(\xbold
(t);\thetabold,
\gbold(t)\bigr)
\end{equation}
in which $\gbold(t)$ can modify $\fbold$ more generally than by
additive forcing. In particular, we will examine allowing a parameter
of interest to vary over time when doing so has a relevant, mechanistic
interpretation. The calculations in \citet{Hooker08}---based on
first-order Taylor expansions---can be readily extended to test
$\gbold(t) \equiv0$ in this more general model.
This approach can be seen as encompassing the model~(\ref{forcing})
and we will use it throughout the paper.

Our new diagnostic tests provide more information about that nature of
the lack of fit when $\gbold(t)$ is found to be significant. In
particular, three nested possibilities for the properties of $\gbold
(t)$ correspond
to the alternatives listed above for how model~(\ref{ode}) should be
reformulated:
\begin{longlist}
\item[\textit{Case} 1.] Exogenous stochastic perturbations: if $\gbold(t)$ is
independent of $\xbold(t)$, this suggests
that $\gbold(t)$ be modeled as a stochastic process, but that the
functional form of (\ref{ode}) is otherwise reasonable.

\item[\textit{Case} 2.] Misspecification of $\fbold$: this is indicated by
$\gbold(t)$ being at least partly
determined by $\xbold(t)$. This would require $\fbold$ to be revised,
as already discussed in \citet{Hooker08}.

\item[\textit{Case} 3.] Missing state variables: if $\gbold(t)$ depends not
only on $\xbold(t)$ but also on past
values $\gbold(t-\delta)$. These lags serve as surrogates for missing state
variables such as additional species in an ecological model, additional
chemical products in a reaction,
or additional ion channels in a neuron. See Section~\ref{secmissing}
for further details.
\end{longlist}

We can motivate this sequence of tests by supposing the data in fact
come from an ODE of the form
%e5 #&#
%
\begin{eqnarray}\label{eqnallBad}
\frac{d\xbold}{dt} & =& \tilde{\fbold}(\xbold,y),
\nonumber\\[-8pt]\\[-8pt]\nonumber
\frac{dy}{dt} & =& k(\xbold,y),
\end{eqnarray}
in which $y$ represents a possible additional state variable and
$\tilde{\fbold}$ represents the true
law of motion that may differ from the assumed law of motion $\fbold$.
Model (\ref{eqnallBad}) has both of the sources of error that we want
to detect. Case~2 corresponds
to $\tilde{\fbold}$ being a function of only $\xbold$, $\tilde
{\fbold}(\xbold,y) = \tilde{\fbold}(\xbold)$.
We consider the additive form of lack of fit (\ref{forcing}). Then we
can write
\[
\gbold(t) = \tilde{f}\bigl(\xbold(t)\bigr) - f\bigl(\xbold(t),\thetabold
\bigr),
\]
so case~2 implies $\gbold(t)$ can be written as a function of $\xbold
(t)$ only.

In case~3 we have
\[
\gbold(t) = \tilde{f}\bigl(\xbold(t),y(t)\bigr) - f\bigl(\xbold
(t),\thetabold
\bigr),
\]
so the time derivative of $\gbold$ is given by
%e6 #&#
%
\begin{eqnarray}\label{eqgderiv}
\frac{d\gbold(t)}{dt} & =& \frac{d\xbold(t)}{dt} \biggl[\frac{d
\tilde{\fbold}(\xbold(t),y(t))}{d \xbold} -
\frac{d \fbold(\xbold
(t),\thetabold)}{d \xbold} \biggr] + \frac{dy(t)}{dt}\frac{d\tilde
{\fbold}(\xbold(t),y(t))}{dy}
\nonumber
\\
& =& \tilde{f}\bigl(\xbold(t),y(t)\bigr) \biggl[\frac{d \tilde{\fbold
}(\xbold
(t),y(t))}{d \xbold} -
\frac{d \fbold(\xbold(t),\thetabold)}{d
\xbold} \biggr]
\\
&&{} + k\bigl(\xbold(t),y(t)\bigr)\frac{d\tilde{\fbold
}(\xbold(t),y(t))}{dy}.\nonumber
\end{eqnarray}
If the map from $(\xbold,y)$ to $(\xbold,\gbold)$ is invertible,
then the expression above implies that
$\frac{dg}{dt} = l(\xbold,\gbold)$ for some function $l$. The
complete dynamical system therefore has\vspace*{1pt} the form
%e7 #&#
%
\begin{eqnarray}\label{eqngState}
\frac{d\xbold}{dt} & =& \tilde{\fbold}(\xbold,\gbold,\thetabold),
\nonumber\\[-8pt]\\[-8pt]\nonumber
\frac{d\gbold}{dt} & =& l(\xbold,\gbold).
\end{eqnarray}
If case~2 holds, the second term in (\ref{eqgderiv}) is zero and the
first term does not depend on $y$, meaning that $d \gbold/dt$ is only
dependent on $\xbold$. This suggests testing for dependence of
$d\gbold/dt$ on $\gbold$, after controlling for $\xbold$, as a way of
distinguishing case~3 from case~2. However, we have found that this
test is statistically less stable than testing whether the
lagged quantity $\gbold(t - \delta)$ helps to predict $\gbold(t)$,
after controlling for $\xbold(t)$. The rationale
for this approach is explained more fully in Section~\ref{secmissing}.

This heuristic can be extended to the model (\ref{forcing2}) if
$\fbold(\xbold(t);\thetabold,\gbold(t))$ is an invertible function
of $\gbold$ for every $\xbold$ and $\thetabold$. However, we note
that if this is not the case---for example, if $\gbold$ is too low
dimensional---we will not be able to completely resolve lack of fit
and this could make a case~2 misspecification appear as case~3.
Apparent case~3 dependence can also result from stochastic fluctuations
if the system evolves probabilistically.

{We also note that (\ref{eqgderiv}) also indicates that there may be
little power to detect case~3 dependence in some systems. In
particular, if $y$ is itself close to being a function of $\xbold$---as
we find to be the case in the chemostat experiments described
below---it will be difficult or impossible to distinguish case~2 from
case~3. }

A system in which parameters are changing systematically (e.g., a
steady upward trend) will also appear as a case~3 type
misspecification, if there is sufficient power to distinguish case~3
from case~2. We believe that this is appropriate. Parameters that are
changing systematically can be considered to have their own dynamics
and are effectively additional state variables.
{Similar comments can be made about systems with stochastic dynamics.}

In this paper, we develop tests to distinguish between each successive
pair of possibilities. These tests need to account for sources of
variation that include resampling methods for the $\ybold_t$ as well
as examining the significance of an appropriate nonparametric regression.
Our methods can be considered as nonlinear continuous-time extensions
of methods to select the number of
lags in linear time-series models and to test between models of
parameter drift; unlike our case, such tests
for linear models can be performed by likelihood ratio tests [see,
e.g., \citet{Hamilton1994}].

To provide a concrete example, we consider a model and data from
experimental population ecology. In the actual experiments
[\citet{Becks10}] algae of the species \textit{Chlamydomonas reinhardtii},
($C$), are grown in a chemostat microcosm which is continuously
supplied with nitrogen-limited medium. These algae are preyed upon by
rotifers of the species \textit{Brachionus calyciflorus}, ($B$),
near-microscopic animals that feed on algae and reproduce asexually
unless at high population density. As a candidate model for this
system, we use a standard predator--prey model from the ecological
literature, the Rosenzweig--MacArthur model:
%e8 #&#
%
\begin{eqnarray}\label{eqRM}
\frac{dC}{dt} & =& r C \biggl(1 - \frac{C}{K_C} \biggr) -
\frac{p G
CB}{K_B + pC},
\nonumber\\[-8pt]\\[-8pt]\nonumber
\frac{dB}{dt} & =& \frac{\chi_B p G CB}{K_B + pC} - \delta B.
\end{eqnarray}
Here $dC/dt$ is the rate of change of the algal population. The first
equation describes this change in terms of logistic growth (because
algae are limited by resource constraints) with maximal growth rate $r$
and carrying capacity $K_C$. This term represents algal birth rate
minus deaths for causes unrelated to predation (in the actual
experiments, washout from the chemostat is the main cause of algal
mortality). The second term represents predation by rotifers. Predation
occurs at maximum rate $G$ but is reduced when algae are scarce, with
$K_B$ representing the algal density $pC$ at which the predation rate
is half of its maximum. The parameter $p$ represents the fraction of
algae available for predation, and is held at $1$ for the moment. Later
we will allow $p$ to vary with time, in providing goodness-of-fit
diagnostics. The equation for the rotifer growth rate $dB/dt$
represents the conversion of consumed algae into rotifers with
conversion rate $\chi_B$, and rotifer mortality $\delta B$ in
proportion to their numbers. Numerically, it is advantageous to
reexpress this system in terms of log variables $\tilde{\xbold} =
(\log C, \log B)$ with differential equation $d \tilde{\xbold}/dt =
\fbold(\exp(\tilde{\xbold});t,\thetabold)/\exp(\tilde{\xbold})$
and we have employed this below.
Note that explicitly modeling washout from the chemostat will be
confounded with parameters $r$, $K_C$, and $\delta$ and we have not
included this in the model.

\begin{figure}

\includegraphics{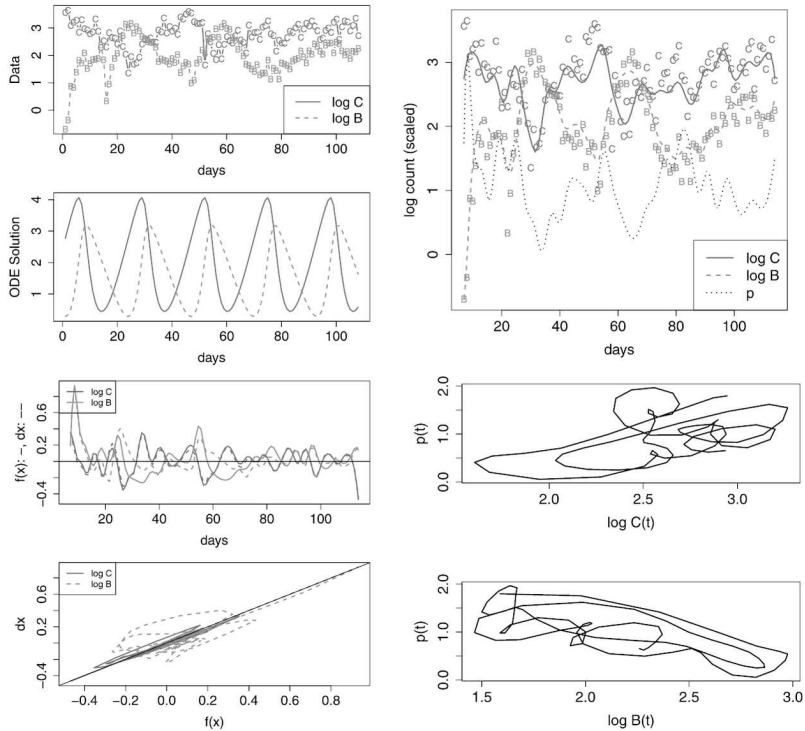}

\caption{Diagnostics for the chemostat data.
{Top left}: time series plot of log data (top) and solution to the Rosenzweig--MacArthur
ODE on log scale with constant $p(t)$ (bottom). These plots allow a
comparison between the qualitative behavior of the observed time series
and of solutions to the ODE model, which produces phase relationships
between $B(t)$ and $C(t)$ different from those in the data.
{Top right}: estimated smooth trajectory $\hat{\xbold}(t)$ and
time-varying $p(t)$. {This allows a comparison of $\hat{\xbold}(t)$
with the data to ensure that our smoothing procedures reflect the data
appropriately}.
{Bottom left}: comparison of $d\xbold/dt$ (dashed
lines) and $\fbold(\xbold;\thetabold,p(t))$ (solid lines) to ensure
that these largely agree after estimating $p(t)$. {Very large
discrepancies relative to the size of $\fbold(\xbold;\thetabold,p(t))$
would indicate that lack of fit has not been adequately
addressed. The lower plot gives $d\xbold/dt$ plotted against $\fbold
(\xbold;\thetabold,p(t))$ for each of $B(t)$ and $C(t)$ to evaluate
the relative size of these departures}.
{Bottom right}: $p(t)$
plotted against $C(t)$ and $B(t)$. {The evident relationships in these
graphs are a visual indicator that $\fbold$ has been incorrectly
specified.}}\label{figChemoDiagnostics}
\end{figure}

The experimental system was sampled once each day, and rofiters and
algae in the sample were counted. Two samples were taken each day, from
the top and bottom of the chemostat, to verify that the system was well
mixed so that spatial variation in population densities does not need
to be considered. The data we analyze are the average of the two daily
samples. Plots of the time series and a fit to these data are given in
the first panel of Figure~\ref{figChemoDiagnostics}; these data come
from \citet{Becks10}, where the experimental methods are
presented in detail.

%f1 #&#
%

A number of features are evident from these plots. Most evidently,
solutions to the ODE have much more regular cycles than the observed
time series. There is also a difference in phase relationships between
the rotifers and algae. In the ODE solutions the rotifer peak is about
$1/4$ cycle period delayed from the algal peak (because rotifer \emph
{population growth rate} peaks when algal density is at a maximum), but
in the observed time series the delay is about $1/2$ the cycle period. A
proposed explanation for this discrepancy [\citet{YJEFH03}] is
that the
algae consist of two subpopulations: one of which does not get predated
but pays a cost in reproducing less efficiently, so that the relative
advantage of each subpopulation is determined by the rate of rotifer
predation. Models incorporating subpopulation structure---hence
expanding the state-vector to $(C_1,C_2,B)$ for two algal
populations---reproduce the
out-of-phase dynamics [\citet{YJEFH03}]. However, this does not rule
out the possibility that the
lack of fit is actually due to misspecifying the functional forms for
the dynamics of the two-dimensional state vector $(C,B)$.

In our examination below, we will allow $p$---the proportion of $C$
that is edible---to vary over time. We examine whether this variation
can be considered random (case~1), is partly determined by $C$ and $B$
(case~2), or also depends on its own past history, indicating a case~3
misspecification. Experimental evidence tells us that the right answer
is case~3 [\citet{YJEFH03}]: when the algal population is homogenous
(all individuals are descended from a single cell), the dynamics are
much more like the predictions of classical predator--prey models such
as (\ref{eqRM}) and do not have a $1/2$-period delay.

To represent time-varying quantities $\gbold(t)$, we employ a basis
expansion, $\gbold(t) = \Psi(t) D$ in which the coefficients $D$ of
the basis function $\Psi(t) = \psi_1(t),\break \ldots,\psi_K(t)$ are
treated as additional parameters to be estimated. Because the addition
$D$ can make the system unidentifiable [e.g., \citet{Hooker08}], we
employ a two-stage estimation procedure, first estimating fixed
parameters $\thetabold$ and then obtaining an estimate for $D$.
Because estimating derivatives by differencing noisy data significantly
increases the noise level and degrades performance, in all of the
methods presented below, $\thetabold$, $\xbold$, and $D$ are
estimated without the need to difference the data.

While the ecological experiment described above provides a useful
motivation, our diagnostics can be employed on a variety of systems. We
explore by simulation the effectiveness of our methods in models for
cardiac rhythms and chaotic dynamics as well as the
Rosenzweig--MacArthur model above. These are investigated both in cases
in which simulated data are generated from an ODE and also when a
stochastic differential equation is used to generate noisy trajectories
which are then observed with noise.

The rest of the paper is structured as follows. Section~\ref
{secestimation} details parameter estimation methods and visual
diagnostics for lack of fit and Sections~\ref{secmisspecify} and \ref
{secmissing} provide testing procedures for misspecification of
$\fbold$ and $\xbold$, respectively. Section~\ref{seclinear}
evaluates these procedures in distinguishing van der Pol and R\"
{o}ssler systems from linear ODEs, while Sections~\ref{secchemo} and
\ref{secheart} investigate these procedures with the nonlinear
Rosenzweig--MacArthur and van der Pol sytems, respectively, along with
applying them to real-world data. We conclude with some speculation
about the power of these tests and further directions to be investigated.

%s2 #&#
\section{Parameter estimation and visual diagnostics} \label{secestimation}

In this section we describe a straightforward method of obtaining
parameter estimates for use in the simulations below. Throughout this
paper we assume that an ordinary differential equation of the form
(\ref{ode}) has been proposed for a system under study in which
$\xbold(t)$ is a $d$-dimensional vector and $\fbold(\xbold;t,\thetabold
)$ takes values in $\real^d$. We further assume that we
have observations $\ybold_i = \xbold(t_i) + \epsilonbold_i$ taken at
times $t_i$ in which each of the state variables is measured with
error. This assumption allows us to use the gradient matching
procedures described below, which we have chosen for the sake of
clarity. However, the tests that we employ can be combined with
alternative parameter estimation methods that do not require
observations of all assumed state variables.

Gradient matching [\citet{Ellner2002}], also referred to as two-stage
least squares in \citet{Wu2012a}, fits parameters of an ODE model
via an initial smoothing step. It proceeds via the following two steps:
\begin{longlist}[2.]
\item[1.] Fit a vector of smooth curves to the data $(t_i,\ybold_i)$ to
obtain estimates $\hat{\xbold}(t)$ of the state variables and their
time derivatives $d \hat{\xbold}/dt$. In our studies below we use
smoothing splines as implemented in the \texttt{fda} package in
\texttt{R} [\citet{RamsayHookerGraves}, see Section~\ref{seclinear} for details],
but alternatives such as local polynomial models [used in \citet
{Ellner2002,Wu2012a}] could also be employed.

\item[2.] Estimate parameters $\thetabold$ by minimizing
$\int[ d \xbold/dt - \fbold(\hat{\xbold}(t);t,\thetabold)
]^2 \,dt$.
\end{longlist}
The first step is implemented in many software packages and the second
may be carried out efficiently with a Gauss--Newton iteration. Note that
if $\fbold$ is linear in its parameters, the second step can be solved
with a simple matrix inversion, a property exploited by \citet
{DattnerKlaasen2013} and which also pertains in our examples.
Importantly, we expect that this procedure will be relatively robust to
model misspecification or disturbances that additively impact $d\xbold
/dt$; this is in contradistinction to fitting solutions to (\ref{ode})
to observed data directly (``trajectory matching'') where local
disturbances of $d \xbold/dt$ can persist in deviations from the unperturbed
solutions for a long time. This means that we expect to be able to
better focus on sources of lack of fit. However,
our tests described below can also be applied using trajectory matching
as a parameter estimation method.

The gradient matching procedure can be readily extended to higher-order
systems. In Section~\ref{secheart} we employ a second-order
representation of the van der Pol equation in one state variable. Here
step~2 is modified to fit the estimated second derivative of $\xbold
(t)$ to a function of its values and its first derivative.

Gradient matching, while simple to implement and present, is limited in
its applicability. Most importantly, it cannot be applied to systems in
which some state variables are not directly measured. It also
introduces bias when there are either relatively few observations or
substantial observation noise. Generalized profiling, introduced in
\citet{RamsayDE}, avoids both these complications by using the
ODE model to improve the smooth in the first step. We have used
generalized profiling with the chemostat example in Section~\ref
{secchemo} and provided a description of these methods in the
supplementary material [\citet{HEsupp}] along with a further set
of simulations.

In our methods we first estimate $\hat{\thetabold}$ in
%\sout{the manner above}
step~2 above with
%\sout{$\gbold(t) \equiv1$}
$\gbold(t) \equiv0$. In order to estimate $\gbold$, we represent it
by another basis expansion: $\gbold(t) = \Psi(t) D$. The coefficients
$D$ are now fit with $\hat{\thetabold}$ held fixed by minimizing the
gradient matching objective: $\int[ d \xbold/dt - \fbold(\hat
{\xbold}(t);t,\thetabold,\Phi(t) D) ]^2 \,dt$.
This two-stage estimation procedure is carried
out to ensure
the identifiability of parameters. Note that
$D$ is estimated within the gradient matching methodology so that the
estimate $\xbold(t)$ will not correspond to an exact ODE solution.

We can now employ the estimate $\hat{\gbold}(t) = \Psi(t) \hat{D}$
to visually examine lack of fit.
First, examining the discrepancy between $d \hat{\xbold}/dt$ and
$\fbold(\hat{\xbold};t,\thetabold,\gbold(t))$ provides a visual
diagnostic of whether time-varying parameters can account for lack of
fit. The procedures we develop here are only appropriate when this is
true, because they presume
that some function $\gbold(t)$ exists that brings the model into line
with the data. If so, we can first
test whether $\hat{\gbold}(t)$ differs from being constant using the
methods in \citet{Hooker08}.
Assuming it does (as we do here), we can then plot $\hat{\gbold}(t)$
versus $\hat{\xbold}(t)$ to look for consistent relationships that
may indicate misspecification of the form of $\fbold$.

These visual diagnostics are demonstrated in Figure~\ref
{figChemoDiagnostics}. The two panels at the top left show the data and
a solution of the proposed Rosenzweig--MacArthur ODE model. The top
right panel shows the smooth curves fitted to
$C(t)$ and $B(t)$ in the first step of gradient matching, and the
estimated $\gbold(t)=p(t)$.
$p(t)$ appears to bear some relationship to both $B(t)$ and $C(t)$
(bottom right panels).
The bottom left panels show that $d\xbold/dt$ and $\fbold(\xbold
;\thetabold,\gbold)$ are fairly similar (i.e., their values lie
near the 1:1 line in the bottom panel), but there remains some
additional departure. This is because Rosenzweig--MacArthur
is an ``off the shelf'' predator--prey model which is not
mechanistically right for the chemostat system.

% {\sout{Our approach does not require us to employ the profiling
%method, however. In Section \ref{secheart} we resort to gradient
%matching} \citep{Ellner2002} \sout{for the sake of computational
%efficiency when working with a frequently-sampled system. This relies
%on smoothing the data to obtain a nonparametric estimate $\hat{
%\xbold}$ of the unobserved smooth trajectory, and then matching $d
%\hat{\xbold}/dt$ to $\fbold(\hat{\xbold};t,\thetabold)$. In the model
%in Section \ref{secheart} our ODE is given in terms of second
%derivatives for a one-dimensional system. This approach can be carried
%out much more quickly than the profiling method -- we used it in other
%systems to obtain initial parameter estimates -- and the same
%calculations we develop below can be employed here as well.}}

%s3 #&#
\section{Tests for dependence between $\gbold(t)$ and $\xbold(t)$}\label{secmisspecify}

{%\sout{Assume now}
For this paper we assume} that $\gbold(t)$ has been shown to differ
from zero,
hence, the ODE mode (\ref{ode}) is misspecified.
We next want to distinguish between the three alternative forms of
misspecification listed in the \hyperref[sec1]{Introduction}.
The first step is to distinguish between alternatives 1 and 2 by asking
whether $\gbold(t)$ has a consistent relationship
with $\xbold(t)$. If so, this indicates that the functional form of
$\fbold$ has been misspecified [because replacing
$\gbold(t)$ with a function of $\xbold(t)$ produces a different ODE
model]. The visual diagnostics above can then indicate help to
determine how $\fbold$ should be amended.

To determine whether $\gbold$ depends on $\xbold$, we assume a null
hypothesis in which $\gbold(t)$ follows a smooth, stationary
stochastic process with zero mean. We attempt to distinguish this from
the alternative hypothesis of some dependence
of $\gbold(t)$ on $\xbold(t)$. This alternative still allows for
error due to genuine random disturbances,
estimation errors, and other forms of misspecification. We conduct this
test via a block-permutation test, using nonparametric estimates for
the relationship between $\gbold(t)$ and $\xbold(t)$. We also account
for the estimation of $\gbold(t)$ through a residual bootstrap.

Formally, our test can be stated as
\begin{eqnarray}
\mathrm{H}_0\dvtx E\bigl(\gbold(t)\mid\xbold(t)\bigr) \equiv0\quad\mbox{versus}\quad \mathrm{H}_A\dvtx E\bigl(\gbold(t)\mid\xbold(t)\bigr) \equiv
\hbold\bigl(\xbold(t)\bigr)\nonumber
\\
\eqntext{\mbox{for some nonconstant function } \hbold(\cdot),}
\end{eqnarray}
where $\hbold$ is assumed to be a sufficiently smooth function
that nonparametric methods can be employed to estimate. This test
could be conducted via a generalized likelihood ratio test [\citet
{FanYao05}], but we must account for the functional nature of $\gbold
(t)$ and $\xbold(t)$ and their estimation.

{To develop a testing procedure for $\mathrm{H}_0$, we first propose a
test statistic given by the form of an $F$-statistic. To calculate
this, we estimate $\hat{\hbold}$ to fit the nonparametric regression model
\[
\hat{\gbold}(t) = \hbold\bigl(\hat{\xbold}(t)\bigr) + \epsilonbold(t).
\]
$\hat{\hbold}$ can be obtained by estimating values of $\hat{\gbold
}$ at a dense set of time points $t_1,\ldots,t_K$,
and then applying any smoothing method that minimizes squared error. In
the simulations and examples below we set the
$t_j$ equal to the observation times in the data and estimated $\hbold
$ by smoothing splines using 40 basis
functions with the default settings in the \texttt{mgcv} package in
\texttt{R} [\citet{mgcv}].
However, our methods are not specific to these choices.}

{We now propose the $F$-statistic}
%e9 #&#
%
\begin{equation}
\label{eqFstatistic} F = \frac{ \sklfrac{1}{K} \sum_{i=1}^K \llVert \hat
{\hbold
}(\hat{\xbold}(t_i)) - \sklfrac{1}{K} \sum_{j=1}^K \hat{\hbold}(\hat
{\xbold}(t_j)) \rrVert ^2 }{ \sklfrac{1}{K} \sum_{i=1}^K \llVert \hat
{\gbold}(t_i) - \hat{\hbold}(\hat{\xbold}(t_i))
\rrVert ^2}
\end{equation}
{as a measure of the strength of association between $\hat{\gbold
}(t_i)$ and $\hat{\xbold}(t_i)$.} {$F$ is analogous
to the standard $F$-statistic for one-way ANOVA, with $\hat\xbold$
values regarded as ``treatment'' levels.}
{Alternative measures such as mutual information could also be
employed. We have chosen the $F$-statistic for its familiarity in
statistical practice and because it can be readily extended to tests
for missing state variables in Section~\ref{secmissing}.}

{We now need to compare $F$ to its distribution if $\mathrm{H}_0$ were
true. We develop this distribution via a two-stage resampling method.
For a fixed $\hat{\gbold}$ and $\hat{\xbold}$, a null distribution
for $F$ can be obtained by a permutation test: permute the values of
$\hat{\gbold}(t_i)$ relative to $\hat{\xbold}(t_i)$ so that any
relationship between $\gbold$ and $\xbold$ is destroyed, re-estimate
$\hbold(\xbold)$, and re-calculate the $F$-statistic. Because of the
continuity of $\hat{\gbold}(t)$, the values of $\hat{\gbold}(t_i)$
exhibit serial dependence over short time intervals, and we therefore
permute these values in blocks.
In addition, we must also account for the variability in the estimates
of $\hat{\gbold}$ and $\hat{\xbold}$. This is done via a residual
bootstrap, and the block-permutation test is conducted within each
bootstrap. This procedure is sketched below, with specific details following:}
\begin{enumerate}[3.]
\item[1.] Estimate $\hat{\xbold}$, $\hat{\thetabold}$, and $\hat
{\gbold}$ from the data.

\item[2.] Estimate $\hat{\hbold}$ to predict $\hat{\gbold}$ from $\hat
{\xbold}$, by smoothing the values
$(\xbold(t_j),\gbold(t_j))_{t=1}^K$. Use the fitted smooth to
calculate $\hbold(t_j)$ values and the $F$-statistic in (\ref{eqFstatistic}).

\item[3.] Evaluate a null distribution for $F$ by a residual bootstrap.
Loop over 1 to~$B_1$:
\begin{longlist}[(a)]
\item[(a)] Create new data by resampling the residuals $\epsilonbold_i =
\ybold_i - \xbold(t_i)$ to create new data
$\ybold_i^b = \xbold(t_i) + \epsilonbold^b$ where the superscript
$b$ indicates a resampled quantity.

\item[(b)] Estimate $\hat{\xbold}^b$, $\hat{\thetabold}^b$, and $\hat
{\gbold}^b$ using the bootstrap data. \label{reboot}

\item[(c)] Estimate $\hat{\hbold}^b$ to predict $\hat{\gbold}^b$ from
$\hat{\xbold}^b$ and calculate the $F$-statistic $F_{0b}$ from~(\ref{eqFstatistic}).

\item[(d)] (Permutation test): loop over $k = 1,\ldots,B_2$:
\begin{enumerate}[(ii)]
\item[(i)] Permute blocks of the vector $\hat{\gbold}^{b}(t_1),\ldots,\hat
{\gbold}^{b}(t_K)$ to create new
values $\hat{\gbold}_1^{kb},\ldots,\hat{\gbold}_K^{kb}$.\vspace*{1pt}

\item[(ii)] Estimate $\hat{\hbold}^{kb}$ to predict the permuted $\hat
{\gbold}^{kb}$ from the $\hat{\xbold}^b$ and calculate the
$F$-statistic $F_{kb}$.
\end{enumerate}

\item[(e)] Measure the significance of $F_{0b}$ by evaluating its $p$-value
relative to the permutation distribution:
\[
p_b = \frac{1}{B_1} \sum_{k=1}^{B_2}
I(F_{0b} > F_{kb}).
\]
\end{longlist}

\item[4.] Assess the significance of the test by rejecting H$_0$ if the
average bootstrap \mbox{$p$-}value is less than $\alpha$: $\sum_b p_p/B_1 <
\alpha$.
\end{enumerate}

{We now elaborate on some of these steps to provide detail. In reverse order:}
\begin{longlist}
\item[\textit{Step} 4] rejects based on an average of
$p$-values. This approach is also taken for tests based on random
projections [\citet{Srivastava2014}]. Under the null, the $p_b$ should
have a uniform distribution. Their average is thus not uniform---it
should be more concentrated around $1/2$. Since the $p_b$ are not
plausibly independent, we cannot derive a null distribution for their
average, and rejecting based on the original significance threshhold is
at least conservative.

\item[\textit{Step} 3(d)(i).] We employ blocks larger than the support
of the basis functions $\Psi(t)$, so that the permutation does not
remove the dependence among close-in-time $\tilde{\gbold}^b$ values
due to the basis function representation. We also remove one half block
at the beginning and end of time points, to avoid edge effects in
estimating~$\gbold$.

\item[\textit{Step} 3(b)] is easily computed when parameters are
estimated by gradient matching, particularly
when $\fbold(\xbold;t,\thetabold)$ is linear in $\thetabold$.
However, this step can be computationally
demanding for profiling methods. For this case, in the supplementary
material [\citet{HEsupp}] we provide a one-step bootstrap based
on a Taylor series expansion.
\end{longlist}

%s4 #&#
\section{Tests for missing dynamical variables} \label{secmissing}

In addition to misspecifying the parametric form of $\fbold$, in
dynamical systems
the proposed model can also misspecify $\xbold$ by omitting important
components of a system.
One example of this is the presence of two visually indistinguishable
subpopulations of algae in the
chemostat system described in the \hyperref[sec1]{Introduction}. Another occurs in
neural dynamics
in which the voltage across the neuron cell membrane is governed by
multiple ion channels
[e.g., \citet{TienGuckenheimer08}, and see \citet{Wilson99}
for an
overview]. Not all of the known channels
are always necessary to describe the dynamics of a single neuron, so
models often focus on a subset of channels,
and lack of fit may result when too few channels are included in a
model. Similar situations can arise
in modeling chemical reactions or pharmacokinetics, if a model omits
some reactions or reaction products.

In this section we assume that a model of the form (\ref{ode}) has
been proposed, but the data actually correspond
to a model of the form
\begin{eqnarray*}
\frac{d\xbold}{dt} & =& \tilde{\fbold}(\xbold,y,\thetabold),
\\
\frac{dy}{dt} & =& k(\xbold,y).
\end{eqnarray*}
To determine whether the proposed model is misspecified in this way,
we seek to {evaluate} evidence that the estimated forcing function
$\gbold(t)$ has additional
internal dynamics that are not accounted for by a functional dependence
of $\gbold$ on $\xbold$.

As we observed above [see equation (\ref{eqngState})], the
difference between this kind of misspecification and the case~2
misspecification considered in the last section
is that how $g(t)$ changes over time depends on $g$ itself, not just on
the putative
state vector $\xbold$. However, we do not directly test for dependence
of $dg/dt$ on $g$.
Instead, motivated by the literature on {\em attractor reconstruction}
[see \citet{abarbanal,KantzSchreiber05} for an overview] that has
developed around the Takens embedding
theorem [\citet{Takens81}], we instead test for dependence of
$g(t)$ on
$g(t-\delta)$.
The methods in this literature predominantly test for dependence on
time-lagged state variables
rather than derivatives because the results are generally more stable
[e.g., \citet{KantzSchreiber05}].
Our experience is in line with this---estimated derivatives were more
noisy, and
our use of a basis expansion creates an unavoidable relationship
between $g$ and $dg/dt$.
As a result, using derivatives instead of time-lagged variables
decreased the power of our tests.
The theorem's underlying attractor reconstruction does not necessarily
hold in stochastic systems or
systems far away from their limiting behavior [although see \citet
{forcedtakens} for extensions].
But for our purposes this is not important. Testing for dependence of
$g(t)$ on $g(t-\delta)$ in addition to $\xbold$
is simply a stable method for seeking evidence that $g$ is a
dynamically evolving state variable whose present state
depends on its past. In contrast, if $g(t)$ is just a function of
$\xbold(t)$, past values of $g$ provide no additional
information about its present value. This qualitative distinction and
the tests we now propose do not depend on the
existence of the transform $l$ or on its invertibility.

The use of a basis expansion induces a relationship between $\gbold
(t)$ and $\gbold(s)$ when $|s-t|$ is small. We
therefore choose $\delta$ to be larger than the support of the
\mbox{B-}spline basis used to estimate $\gbold(t)$, specifically
twice the block length employed in the block permutation test. With
this in mind, we
can state our test of missing components explicitly as
\[
\mathrm{H}_0\dvtx E g_i(t) \equiv\hbold_0\bigl(
\xbold(t)\bigr)\quad\mbox{versus}\quad \mathrm{H}_A\dvtx E g_i(t)
\equiv\hbold_1\bigl(\xbold(t),g_i(t-\delta)\bigr).
\]
We will approach this test using the same ideas as in the previous
section. {To do so, we construct smooths $\hat{\hbold}_0$ and $\hat
{\hbold}_1$ corresponding to the two hypotheses above and calculate an
$F$-statistic for the difference in predictions between these.
Specifically, we define
%e10 #&#
%
\begin{equation}
\label{eqFstatistic2} F = \frac{ \sklfrac{1}{K} \sum_{i=1}^K \llVert \hat
{\hbold
}_1(\hat{\xbold}(t_i),\hat{\gbold}(t_i-\delta)) - \hat{\hbold
}_0(\hat{\xbold}(t_i)) \rrVert ^2 }{ \sklfrac{1}{K} \sum_{i=1}^k \llVert
\hat{\gbold}(t_i) - \hat{\hbold}_1(\hat
{\xbold}(t_i),\hat{\gbold}(t_i-\delta)) \rrVert^2}.
\end{equation}
For this we again use the functions in the \texttt{mgcv} package, but
any smoothing method could be employed.
We also need to modify the permutation test, which we do by permuting
the residuals from the null model $\etabold(t) = \gbold(t) - \hat
{\hbold}_0(\xbold(t))$ in blocks to create a data set in which H$_0$
is true. }

To carry this out, we proceed following the procedure given in
Section~\ref{secmisspecify}, modifying only the following steps:
\begin{enumerate}[3(c)]
\item[3(c)] Estimate $\hat{\hbold}_0^b$ to predict $\hat
{\gbold}^b$ from $\hat{\xbold}^b$ and $\hat{\hbold}_1^b$ to
predict $\hat{\gbold}^b$ from both $\hat{\xbold}^b$ and $\hat
{\gbold}^b(t-\delta)$ and calculate the $F$-statistic $F_{0b}$ from
(\ref{eqFstatistic2}).

\item[3(d)] (Permutation test): loop over $k = 1,\ldots,B_2$:
\begin{longlist}[(a)]
\item[(a)] Permute blocks of the residual vector $\etabold^b(t_i) = \gbold
(t_i) - \hat{\hbold}_0(\xbold(t_i))$ and add these to predictions to
create $\hat{\gbold}^{kb}_j = \hat{\hbold}_0^b( \xbold(t_j) ) +
\etabold^{kb}(t_j)$.
$\hat{\gbold}_1^{kb},\break\ldots,\hat{\gbold}_K^{kb}$.

\item[(b)] Estimate $\hat{\hbold}_0^{kb}$ to predict $\hat{\gbold}^{kb}$
from $\hat{\xbold}^b$ and $\hat{\hbold}_1^{kb}$ to predict $\hat
{\gbold}^{kb}$ from both $\hat{\xbold}^b$ and $\hat{\gbold
}^{kb}(t-\delta)$ and calculate the $F$-statistic $F_{0b}$ from (\ref{eqFstatistic2}).
\end{longlist}
\end{enumerate}
This test can thus be run alongside the test in Section~\ref{secmisspecify}.

%s5 #&#
\section{Simulation example: Linear systems versus van der Pol and R\"{o}ssler systems}
\label{seclinear}

We have a set of four nested hypotheses concerning the misspecification
of the system, which we can write as:
\begin{longlist}[H0.]
\item[H0.] $\gbold(t) \equiv0$,

\item[H1.] $E[ \gbold(t)|\xbold(t), \gbold(t-\delta)] \equiv0$,

\item[H2.] $E [\gbold(t)|\xbold(t),\gbold(t-\delta)] = h(\xbold (t))$,

\item[H3.] $E [\gbold(t)|\xbold(t),\gbold(t-\delta)] = l(\xbold (t),\gbold(t-\delta))$.
\end{longlist}
In the previous sections we have proposed tests to distinguish H2 from
H1 and H3 from H2.
\citet{Hooker08} presents methods to distinguish H1 from H0.
We now examine the performance of these tests using simulations
and real data.

In our first experiment the proposed model is the 2-dimensional linear system
\begin{eqnarray*}
\frac{dx_1}{dt} & =& a_{11}x_1 + a_{12},
x_2,
\\
\frac{dx_2}{dt} & =& a_{21} x_1 + a_{22}
x_2
\end{eqnarray*}
with the $a_{ij}$ as unknown parameters. We examine three
data-generating models:
\begin{longlist}[2.]
\item[1.] Circular motion, which corresponds to the linear model with
$(a_{11},a_{12},\break a_{21},  a_{22}) = (0,-1,1,0)$.
In this case H0 is true, because the model is correctly specified.

\item[2.] The van der Pol oscillator [\citet{vdp27}]:
\begin{eqnarray*}
\frac{dx_1}{dt} & =& a x_2,
\\
\frac{dx_2}{dt} & =& b \biggl(x_2 - x_1 -
\frac{x_2^3}{3} \biggr),
\end{eqnarray*}
in which misspecification appears as an additive term in the equation
for $x_2$. {In this case H2 is true}.
We take $(a,b) = (0.25,4)$.

\item[3.] The R\"{o}ssler system [\citet{Rossler76}]:
\begin{eqnarray*}
\frac{dx_1}{dt} & =& - x_2 - z,
\\
\frac{dx_2}{dt} & =& x_1 + a x_2,
\\
\frac{dz}{dt} & =& b + z(x-c).
\end{eqnarray*}
In this case the true state vector includes a third variable, so H3 is true.
We take $(a,b,c) = (0.2,0.2,3)$, and we also consider values $(a,b,c) =
(0.2,0.2,5.7)$,
parameter values classically chosen to produce chaotic dynamics.
\end{longlist}

For each of these we will examine data generated from the differential
equation and data
from a stochastic differential equation with additive noise
corresponding to
%e11 #&#
%
\begin{equation}
\label{eqsde} d \xbold= \fbold(\xbold,\thetabold) \,dt + \sigma \,d\Wbold,
\end{equation}
where $\Wbold$ is a multivariate Wiener process with independent components.
{For the systems above,} we took $\sigma^2 = 0.01$ for the linear and
van der
Pol models and $\sigma^2 = 0.004$ for the R\"{o}ssler system. These
choices gave us a range of stochastic variabilities without making the
nonlinear systems diverge to infinity. {For the R\"{o}ssler system with
chaotic parameter values,
the stochastic system exhibits noticeably shorter-period oscillations;
we therefore sped up the
ODE experiments by multiplying the right-hand side of this system by a
factor of 2,
which gave periods similar to the stochastic version. }

{For each of these systems, we generated a set of observations by
adding Gaussian noise to the state of the system:
\[
\ybold_i = \xbold(t_i) + \epsilonbold_i,
\]
where the $t_i$ are taken to be 440 equally spaced time points from
$t=0$ to $t=55$ and the $\epsilonbold_i$ are independent Gaussians
with variances 0.25, 0.001, and 0.01 for the linear, van der Pol, and
R\"{o}ssler systems, respectively. For the R\"{o}ssler system, only
$x_1$ and $x_2$ were observed. In each case we estimated an empirical
forcing function $g(t)$ that was added to the second state
variable $x_2$. We used cubic B-splines with a second-derivative
penalty to generate $\hat{\xbold}$ based on knots every 0.25 time
intervals with penalty parameter 0.01; some undersmoothing at this step
is recommended to reduce bias [\citet{Ellner2002}]. $g(t)$ was
represented by a cubic B-spline with knots at integer time intervals
from 0 to 55. Each simulation was repeated 200 times.}

%f2 #&#
%
\begin{figure}[b]

\includegraphics{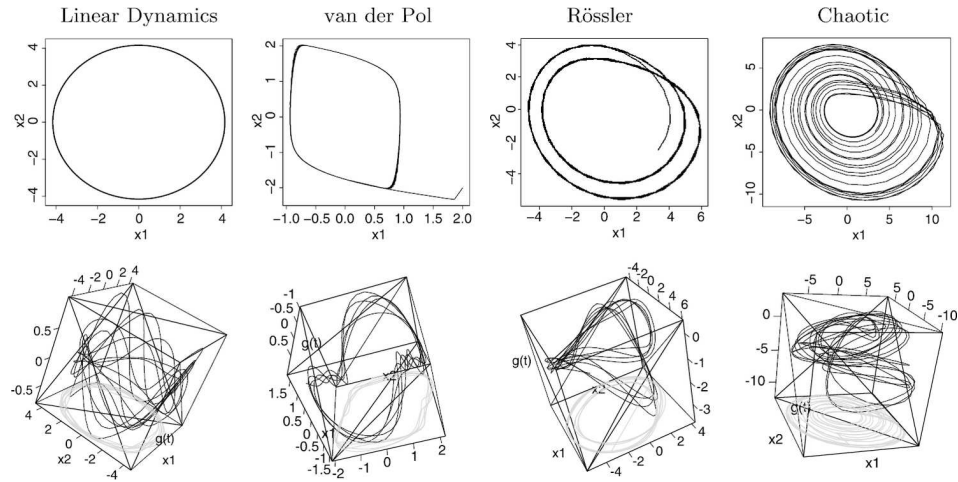}

\caption{Diagnosing lack of fit for the linear model fitted to data
generated by the linear,
van der Pol, R\"{o}ssler, and chaotic R\"{o}ssler systems.
Top row: phase plane plots of the state variables $x_1,x_2$ in the ODE
(deterministic) model
that were sampled to create the data series.
Bottom: diagnostic plots of $\hat\gbold(t)$ plotted against $x_1(t)$
and $x_2(t)$. Black curves
are the three-dimensional trajectories of the SDE, and the grey curves
are their projections onto the
$(x_1,x_2)$ plane. {A clear functional relationship is especially
visible for the van der Pol example,
suggesting correctly that \textup{H2} is true in this case.}}\label{figlinsims}
\end{figure}

{Visual diagnostics for lack of fit are given in Figure~\ref
{figlinsims}, which shows
three-dimensional representations of the empirical relationship between
$g(t)$, $x_1(t)$, and $x_2(t)$.
For the linear data we see no relationship, correctly supporting H0.
For the van der Pol data, we see a clear functional
depence of $g$ on $(x_1,x_2)$, correctly supporting H2. For the R\"
{o}ssler and Chaotic data, there is no single-valued functional
relationship. Rather, the plots suggest trajectories of a three- (or
more) dimensional dynamical system, which correctly supports
H3.}

{The power of our proposed tests for each of these systems is given in
Table~\ref{tablinsims}. For the linear system, the formal
tests correctly do not detect any lack of fit, and for the van der Pol
system the tests correctly reject H1 against H2 with
high power, but do not reject H2 against H3. For the R\"{o}ssler and
Chaotic, H1 and H2 should both be rejected, but this
does not always occur with high power. In these systems, the
unequivocal evidence for presence of an unmeasured third state variable is
that trajectories in the $(x_1,x_2)$ plane cross each other, which
cannot happen in any ODE with $(x_1,x_2)$ as the only state variables.
In these simulations, such crossings only occur in a limited region of
the two-dimensional state space, and this may account for
the reduction in power.}

%t1 #&#
%
\begin{table}
\tabcolsep=0pt
\caption{Power of goodness-of-fit test for case~2 (misspecification of
$\fbold$) and case~3 (missing components in $\xbold$) for
data generated by the linear, van der Pol and R\"{o}ssler ODE and SDE
models following parameter estimation by gradient matching. These were
estimated from 200 simulations for each model as described in
Section~\protect\ref{seclinear}}\label{tablinsims}
\begin{tabular*}{\tablewidth}{@{\extracolsep{\fill}}@{}lccccc@{}}
\hline
& & \textbf{Linear dynamics} & \textbf{van der Pol} & \textbf{R\"{o}ssler} & \textbf{Chaotic}\\
\hline
ODE model & Case 2 (H2 v H1) test & 0.06\phantom{0} & 1 & 1\phantom{.000} & 1\phantom{.00}\\
& Case 3 (H3 v H2) test & 0.005 & 0 & 0.48\phantom{0} & 1\phantom{.00}
\\[3pt]
SDE model & Case 2 (H2 v H1) test & 0.01\phantom{0} & 1 & 1\phantom{.000} & 0.91 \\
& Case 3 (H3 v H2) test & 0.005 & 0 & 0.915 & 0.68 \\
\hline
\end{tabular*}
\end{table}

{Overall, our tests are somewhat conservative for these test cases. We
would expect that the power of our tests would increase with longer
time intervals and more frequent data, but would likely decrease as the
dimension of the systems under study increases. However, our tests do
have reasonable power to detect relevant types of misspecification in
these models.}

%s6 #&#
\section{Example: Chemostat models} \label{secchemo}

In this section we present the application of these tests to assess
evidence for evolution in the chemostat models described in the
\hyperref[sec1]{Introduction} and shown in Figure~\ref{figChemoDiagnostics}, with the
Rosenzweig--MacArthur model~(\ref{eqRM}) as the proposed model.
{Because of the relative sparsity of the experimental data, we
estimated model parameters using the profiling methods described in the
supplementary material
[\citet{HEsupp}],
rather than gradient matching as described in Section~\ref
{secestimation}. All other aspects of testing the model remain the same.}

%f3 #&#
%
\begin{figure}

\includegraphics{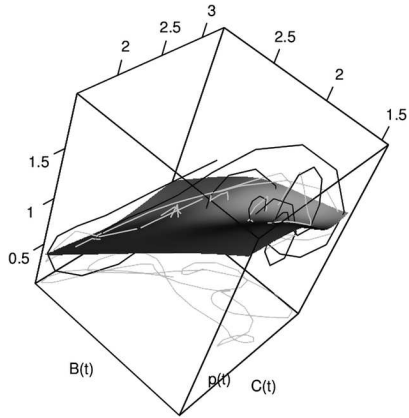}

%Misspecification of $\fbold$ p-value: 0.052 Misspecification of $
%\xbold$ p-value: 0.43 \\
\caption{Visualization of the diagnostic tests for the
Rosenzweig--MacArthur model applied to the
chemostat data. Surface indicates predictions of $p(t)$ based only on
$(C(t),B(t))$; Dark lines are
$p(t)$ plotted against $(C(t),B(t))$; Light lines are predictions of
$p(t)$ based also on $p(t-\delta)$.}
\label{figChemoDiagnosis}
\end{figure}

Figure~\ref{figChemoDiagnosis} presents the estimated time-varying
trait $p(t)$ plotted against the estimated $C(t)$ and $B(t)$
{(represented by a cubic B-spline basis with knots every 0.5 days)},
along with a surface representing the smooth of this relationship, and
predictions from a model that also includes $p(t - \delta)$ {where
$p(t)$ was parameterized by a cubic B-spline basis with knots every 3
days}. There is apparent misspecification of $\fbold$ (H2 against H1),
although the $p$-value for this (0.052) falls short of the traditional
threshold for significance. There is insufficient evidence ($p = 0.45$)
that the state variable is missing a component (H3 against H2), which
could be produced by an additional algal subpopulation.

\begin{figure}

\includegraphics{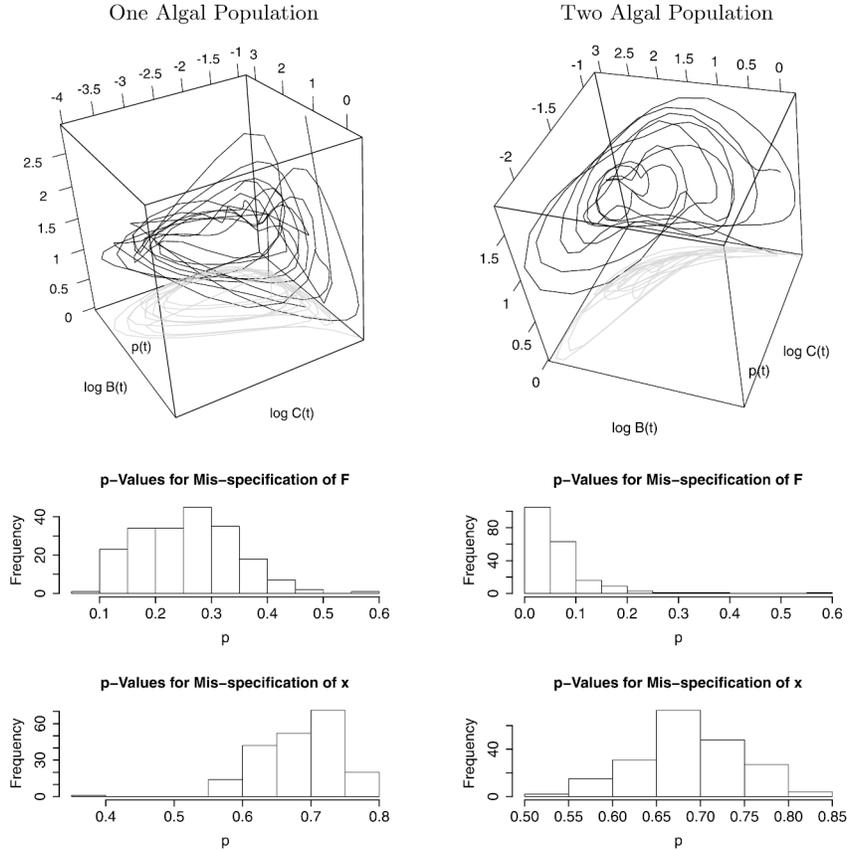}

\caption{Top row: example diagnostic plots for a Rosenzweig--MacArthur
model with only one algal population
fitted to data from a chemostat system model with either one algal
population (left) or two algal populations (right). Bottom: histograms
of $p$-values tests for misspecification of the dynamics $\fbold$ (top)
and misspecification of the state vector (bottom), based on 200 simulations.}\label{figchemosim}
\end{figure}

However, these results do not warrant the conclusion that evolution
does not occur in this system, indeed, additional experiments proved
that it does [\citet{YJEFH03}]. The tests rely on the system producing
behaviors in which this type of dependence can be readily uncovered.
For this system, the power to detect such lack of fit is very low. To
demonstrate this, we conducted a simulation study based on two
plausible, more complex, stochastic models for the rotifer-algae
system. Details of these models are in the supplementary material
[\citet{HEsupp}]. The salient distinction between the two models is
that one of them includes two populations of algae, while the other
does not. We again simulated 200 data sets from each and conducted the
proposed tests. Figure~\ref{figchemosim} presents histograms of the
\mbox{$p$-}values for each test along with example plots relating $p(t)$ to
$B(t)$ and $C(t)$ in each model. Here we see that misspecification of
$\fbold$ is detectable ($p$-value $< 0.05$ in 53\% of the data sets)
in the two-algal population model, but the test for missing state
variables has very little power (0 out of 200 in both models). The
diagnostic plots of Figure~\ref{figchemosim} are helpful in
explaining why this is the case; the grey lines produce the
design of covariates values for the case~2 regression of $p(t)$ on
$(C(t),B(t))$.
Here we see that while the model that incorporates multiple algal types
produces cycles which are much more elongated, the cycles still do not
cross (as they do in the R\"{o}ssler system in Figure~\ref
{figlinsims}). This means that an appropriate nonlinear dependence
of $p(t)$ on $(C(t),B(t))$ can capture all of the signal in this
relationship, so adding $p(t-\delta)$ as a covariate will not improve
predictive performance.

%f4 #&#
%

This example provides the important practical lesson that detection of
missing state variables requires
the system to behave in ways that cannot be replicated by \textit{any
dynamical model} that uses the current state space.
In this case, there are mechanisms besides algal evolution that can
generate the observed system behavior.
Once the system is close to its stable periodic trajectory, the
relative abundance of the two different
algal types can be predicted from the rotifer abundance and total
algal abundance (as seen in the functional relationships of $p$ with
$B$ and $C$ in the bottom right panels of Figure~\ref{figChemoDiagnostics}).
Inserting this dependence into the rotifer's feeding rate equation
[where $p(t)$ has the largest effect]
produces a two-variable model that can exhibit the kind of antiphase
cycles seen in the
experiment with two algal subpopulations. We hypothesize that this
modification to the predator's feeding rate equation
serves as a proxy for predator age structure, allowing the model to behave
like models that can exhibit the kind of antiphase cycles seen in the
experiment as a result of predator age structure.
Independent experimental evidence tells us predator age structure is
not the mechanism operating in these experiments
[\citet{YJEFH03,hiltunen-etal-2014}], but from the time series
alone it
may not be possible to determine that the actual
mechanism involves additional state variables.

%f5 #&#
%
\begin{figure}[b]

\includegraphics{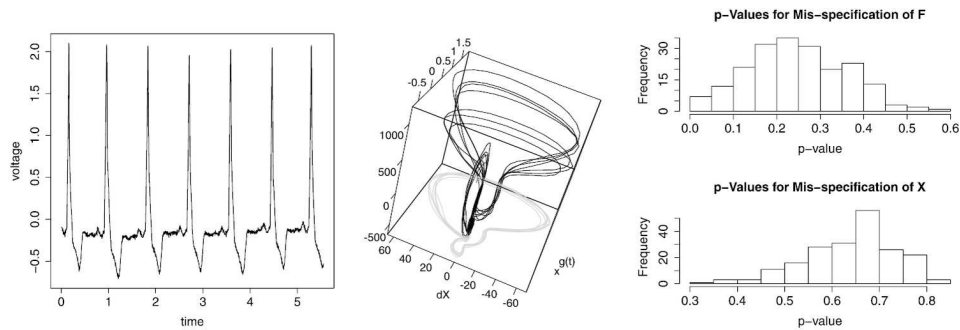}

\caption{Left: electro-cardiogram data. Middle: diagnostic plots for
the van der Pol model indicating both cases~2~and~3
misspecification. Right: histograms of $p$-values for data simulated
from a van der Pol model without misspecification.}\label{figheart}
\end{figure}

We also undertook 200 simulations employing the ODE model (\ref
{eqRM}), transformed to represent $\log C(t)$ and $\log B(t)$, to
generate data along with additive Gaussian errors with variance 0.25.
This provides a means of checking that the nonlinearity of these
equations does not distort our tests. The levels of both tests were
estimated from this simulation at 0, indicating that the test remains
conservative in the presence of nonlinearities.

%s7 #&#
\section{Example: Cardiogram data and the van der Pol system} \label
{secheart}

In this section we present data from electro-cardiogram measurements
obtained from the \mbox{MIT-BIH} Arrhythmia Database [subject~214, \citet{MITBIH00,MoodyMark01}], given in the first plot of
Figure~\ref{figheart}. For these data we employ an alternative formulation of the
van der Pol model studied in Section~\ref{seclinear} that is given as
a second-order differential equation
%e12 #&#
%
\begin{equation}
\label{eqvdp2} \frac{d^2x }{dt^2} = a + b \frac{dx}{dt} + cx +
dx^2 + e x \biggl(\frac{dx}{dt} \biggr)^2.
\end{equation}
The van der Pol model places further restrictions on the parameters
$a$, $b$, $c$, $d$, and~$e$, but we leave these to be estimated independently.
For this system we employ an extension of gradient matching to second-order ODE's by estimating two derivatives: $\hat{x}$, $d \hat{x}/dt$,
and $d^2\hat{x}/dt^2$ {using a cubic B-spline basis with 500 knots
across the time interval}. We then choose parameters to minimize
\[
\int\biggl( \frac{d^2\hat{x}}{dt^2} - a - b \frac{d\hat{x}}{dt} - c\hat
{x} - d
\hat{x}^2 - e \hat{x} \biggl(\frac{d\hat{x}}{dt} \biggr)^2
\biggr)^2 \,dt.
\]
This can be carried out by evaluating the estimated smooth and its
derivatives at a fine grid of time points and then employing linear
regression. Following this, the residuals are smoothed using an
{unpenalized cubic B-spline basis expansion with knots every 0.05
seconds---about 8 observations per knot}---to obtain an estimated
$g(t)$ as a lack of fit forcing function.
The testing procedure proceeds as above with model misspecification
obtained by relating $g(t)$ to $x(t)$ and $dx/dt$, and tests for
missing state variables carried out by testing whether $g(t-\delta)$
provides additional predictive accuracy.

{ A visual display of the analysis for this system is given in
Figure~\ref{figheart}. The middle panel, in particular, plots the
estimated $g(t)$ against $x(t)$ and $dx/dt$. Here we see a consistent
relationship, but also an evident, nearly vertical ``cycle'' that is
preserved across multiple heart beats. This cycle corresponds to the
small, but consistent bump in the left-hand plot just before the main
spike in voltage. It presents a visual indication of missing state
variables, where knowledge of $g(t-\delta)$ can distinguish which part
of the subcycle the system is in. To formally test this conclusion,%
%\sout{In these data }
}
we left off the first and last 100 time points in our testing
procedures, and used blocks of size 50. %\sout{We bootstrapped
%residuals from the smooth to construct each new data set.}
Here both tests returned $p$-values of zero, indicating that both types
of misspecification are present {and confirming our visual impression. %
%\sout{In the second plot in Figure \ref{figheart}, the missing
%dynamic component is only relevant in the ``knot'' observable where
%$g(t)$ undergoes a sub-cycle when plotted against $x(t)$ and $dx/dt$.}
}

To ensure that this effect was not an artifact of the estimation
methodology, we conducted a simulation study employing solutions to
(\ref{eqvdp2}) as the data,
with additive observation noise, so that the fitted model is correctly
specified. Histograms of \mbox{$p$-}values from both tests are given in the
final plot of Figure~\ref{figheart}. Although these are not uniformly
distributed, the level of the test is at least conservative (0.035 for
case~2, 0 for case~3).

%s8 #&#
\section{Conclusions}
This paper represents lack of fit in differential equation models as a
series of nested hypotheses:
\begin{longlist}[2.]
\item[1.] No lack of fit.

\item[2.] Unaccounted-for stochastic variation.

\item[3.] Misspecified right-hand side functions for the differential equation.

\item[4.] Missing or misspecified state variables that describe the system.
\end{longlist}
We presented tests to distinguish the third from the second and the
fourth from the third of these. This nested structure is necessary for
the last two possibilities,
but nesting the second and third is not strictly required. However, we
believe this nesting makes sense in analogy to regression model
diagnostics which include a random error term. {Lack of fit can
alternatively be tested by proposing alternative parametric models and
comparing model likelihoods; to our knowledge, this paper is the first
attempt to produce tests that distinguish between different kinds of
lack of fit
without explicitly modeling them. }

Our tests rely on bootstrap and permutation methodologies in order to
require as few assumptions as possible. This leads to their being
conservative at the null hypothesis; it also makes conducting them
computationally demanding. However, they are still capable of
distinguishing meaningful differences between models, as our
simulations indicate. While our methods are based on explicitly smooth
models of dynamics, we have also demonstrated that these systems work
well with nonsmooth diffusion processes.

The nonparametric nature of these tests can reduce their power.
{Moreover, some systems exhibit dynamics in which detecting a missing
component is fundamentally difficult.} As our ecological example
indicates, genuinely three-dimensional systems can often be represented
as two-dimensional systems, unless they have behavior that cannot be
embedded in two dimensions, {and this confounds the two tests that we
propose. Methods to distinguish which systems will exhibit this type of
confounding are an important direction for future research.} More
powerful tests can be based on specific alternative hypotheses. For
example, the two-algal population
model given in the supplementary material [\citet{HEsupp}] provides
better qualitative agreement with the data than does the elaborated
one-algal model. However, neither model is exactly correct,
and tests to distinguish between them while making few assumptions
about the form of a stochastic model have yet to be developed.

There is also room to design experiments that would yield behavior in
which missing state variables,
such as the second algal population in the chemostat data, is more
readily detected by the tests proposed
here. \citet{HookerLinRogers13} and \citet
{ThorbergssonHooker13} present some experimental design methods for
dynamical systems {in which inputs are perturbed so that observations
yield optimal
information about parameters of interest}. Mork work is needed to adapt
these techniques to our tests. The power of our
test for misspecified state variables also might be higher when several
trajectories have been observed that have
different initial values. The test fails when the trajectory of an
$n$-dimensional system, projected onto
$n-k$ dimensions, can be reproduced or approximated well by the
solution of some $(n-k)$-dimensional
dynamical system. This is especially likely if the observed trajectory
is on or near a low-dimensional
attractor for the dynamics and the dynamics are close to deterministic
because of the Takens Embedding Theorem
[\citet{Takens81}]. A second trajectory, with initial values far from
the attractor, might require a higher-dimensional
system or a different lower-dimensional system to reproduce it, and
these would reveal that the
system is actually higher dimensional.

\begin{supplement}[id=suppA]
\stitle{Supplementary material for ``Goodness of fit in nonlinear dynamics: Misspecified rates or misspecified states?''}
\slink[doi]{10.1214/15-AOAS828SUPP} %[doi,text={...}] - jei reikia
%suskaldyti doi
\sdatatype{.pdf}
\sfilename{aoas828\_supp.pdf}
\sdescription{This appendix provides supporting material which
includes the following: details of the chemostat models used to
generate data for Section~\ref{secchemo} and background material on
the generalized profiling methods of \citet{RamsayDE}, along with
simulation experiments using this method instead of gradient matching.}
\end{supplement}

% imsref loaded by linak, 2015-05-14 10:37:12
%

\printaddresses
\end{document}